\documentclass[conference]{IEEEtran}
\usepackage{amsmath}
\usepackage{amsfonts}
\usepackage{graphicx}
\usepackage{subfigure}
\usepackage{algorithm}
\usepackage{algorithmic}
\usepackage{multirow}
\usepackage{latexsym}
\usepackage{cite}

\ifCLASSINFOpdf
\else
\fi
\begin{document}
%
\title{Hu-Fu: Hardware and Software Collaborative Attack Framework against Neural Networks}

\author{\IEEEauthorblockN{Wenshuo Li, Jincheng Yu, Xuefei Ning, Pengjun Wang, Qi Wei, Yu Wang, Huazhong Yang}
\IEEEauthorblockA{Department of Electronic Engineering, Tsinghua University\\
Beijing National Research Center for Information Science and Technology\\
\{lws17@mails.tsinghua.edu.cn, yu-wang@tsinghua.edu.cn\}}
}


%


\maketitle

\begin{abstract}
Recently, Deep Learning (DL), especially Convolutional Neural Network (CNN), develops rapidly and is applied to many tasks, such as image classification, face recognition, image segmentation, and human detection. Due to its superior performance, DL-based models have a wide range of application in many areas, some of which are extremely safety-critical, e.g.
intelligent surveillance and autonomous driving. Due to the latency and privacy problem of cloud computing, embedded accelerators are popular in these safety-critical areas. However, the robustness of the embedded DL system might be harmed by inserting hardware/software Trojans into the accelerator and the neural network model, since the accelerator and deploy tool (or neural network model) are usually provided by third-party companies. Fortunately, inserting hardware Trojans can only achieve inflexible attack, which means that hardware Trojans can easily break down the whole system or exchange two outputs, but can't make CNN recognize unknown pictures as targets. Though inserting software Trojans has more freedom of attack, it often requires tampering input images, which is not easy for attackers. So, in this paper, we propose a hardware-software collaborative attack framework to inject hidden neural network Trojans, which works as a back-door without requiring manipulating input images and is flexible for different scenarios. We test our attack framework for image classification and face recognition tasks, and get attack success rate of 92.6\% and 100\% on CIFAR10 and YouTube Faces, respectively, while keeping almost the same accuracy as the unattacked model in the normal mode. In addition, we show a specific attack scenario in which a face recognition system is attacked and gives a specific wrong answer.
\end{abstract}

%
\IEEEpeerreviewmaketitle

\section{Introduction}
Deep Learning (DL) has experienced rapid growth. From AlexNet\cite{alexnet} to ResNet\cite{resnet}, the top-5 accuracy of classification task raised from 84.7\% to 96.4\% in Image-Net Large Scale Vision Recognition Challenge (ILSVRC)\cite{ilsvrc}. 
Due to its good performance, deep learning has shown a promising application in many new areas such as intelligent surveillance\cite{intelli-sur}, autonomous driving\cite{auto-drive} and smart home\cite{smart-home}.

Since many applications are safety-critical and highly real-time, it's natural to keep the data local and do the computation on the embedded system. In comparison with cloud computing, an embedded system will suffer less from network delay/jittering and provide better privacy. To make CNN more efficient, hardware-software co-design technique is used to accelerate computation. In terms of hardware design, 
there has been much previous work. Diannao\cite{diannao} gave a design of neural network accelerators and achieved 452 GOP/s performance and 485 mW power consuming. Qiu et al.\cite{go-deeper} presented a software-hardware co-design method to make the computation faster, using SVD and data quantization. As it shows great potential, the industry devotes to product development, such as Google's TPU\cite{tpu} and DeePhi's DPU\cite{dpu}.

In terms of software design, there is also plenty of work. Han et al.\cite{hansong-iclr} introduced Deep Compression to significantly reduce the storage requirement of CNN, which means less energy would be used in data handling. Li et al.\cite{prune-channel} and He et al.\cite{he2017channel} made research on coarse-grained pruning. Yang et al.\cite{mit-prune} presented energy-aware pruning to achieve higher energy efficiency. Fixed-point training technique is also studied a lot. Courbariaux et al.\cite{BNN} gave a way to binarize parameters and achieved less storage and bandwidth consuming. Zhou et al.\cite{do-re-fa} searched different fixed bits and made a comparison. This work achieved better performance while keeping low storage. Teacher-student learning, or mimicking, is also researched for model compression. Ba et al.\cite{teacher-net} 
introduced teacher-student training. An improvement is made in\cite{distillation}, in which distillation is used 
to make student networks easier to learn. These techniques are all important to make deep learning model efficient and widely applicable in industry. DNNDK\cite{dnndk} by DeePhi is a powerful software tool which compresses model before deploying deep learning model on the accelerator.

\begin{table}[hpt]
\centering
\caption{comparison of threat targets}
\label{table:attack-comparison}
\begin{tabular}{c|c|c|c|c}
\hline
                                                                   & \begin{tabular}[c]{@{}c@{}}training\\ data\end{tabular} & \begin{tabular}[c]{@{}c@{}}input\\ image\end{tabular} & \begin{tabular}[c]{@{}c@{}}model\\ parameters\end{tabular} & \begin{tabular}[c]{@{}c@{}}hardware\\ architecture\end{tabular} \\ \hline
\begin{tabular}[c]{@{}c@{}}adversary examples\\\cite{ae1,ae2,ae3,ae4,ae5,phy1,phy2}\end{tabular}       & no                                                      & yes                                                   & no                                                         & no                                                              \\ \hline
\begin{tabular}[c]{@{}c@{}}data poisoning\\\cite{data-poison1,data-poison2}\end{tabular}           & yes                                                     & no                                                    & yes                                                         & no                                                              \\ \hline
\begin{tabular}[c]{@{}c@{}}neural network trojans\\\cite{nnt1, nnt2}\end{tabular} & yes                                                     & yes                                                   & yes                                                        & no                                                              \\ \hline
\textbf{proposed}                                                  & \textbf{no}                                             & \textbf{no}                                           & \textbf{yes}                                               & \textbf{yes}                                                    \\ \hline
\end{tabular}
\end{table}

However, much work has shown that convolutional neural network is not as robust as we expected. We categorize them into different types by threat targets, shown in table~\ref{table:attack-comparison}. \cite{ae1, ae2, ae3, ae4, ae5} show that CNN is easily confused by imperceptible adversarial perturbation on input test images.
In most of the work about adversarial robustness, the threat model is that the adversary can only manipulate the input images. 
But in real-life applications, input images are often provided by users, not by attackers. So this kind of attack is not easy to achieve. Some work has been made to bring these attacks into the physical world, such as \cite{phy1, phy2}. Another type of attack is called data poisoning \cite{data-poison1, data-poison2}. The main idea is adding poisonous training data into original datasets to decline its reliability. A variant of data poisoning is neural network Trojans \cite{nnt1, nnt2}, which often insert designed patterns into original training dataset to make CNN give a specific wrong answer when the test image contains the pattern. Neural network Trojans also tamper input images. Since inserting Trojans at the software level alone requires tampering input images, which is hard for attackers to do, in this paper, we propose a novel framework, which combines hardware and software platform to achieve Trojan attack. This paper makes the following contributions.

\begin{itemize}
\item We define a threat model of neural network attack. Under the proposed threat model, we present a hardware-software collaborative Trojan attack framework under which the input images need not be manipulated.
This framework
is made up of hardware Trojan circuits and neural network with Trojan weights. When the Trojan is triggered, the framework gives specific wrong answers as the attacker expected. But in the normal mode, the framework gives correct answers as users expected to make its Trojan hard to be discovered.
\item Inspired by DSD\cite{dsd}, we propose a training process to insert Trojans without influencing the original accuracy. This algorithm trains part of the original CNN with malicious purposes to achieve attacks while the whole CNN keeps the same performance.
\item We test our attack framework for image classification and face recognition tasks. We achieve attack success rate of 92.6\% and 100\% on CIFAR10 and YouTube Faces Database, respectively. We show a specific attack scenario in which a face recognition system is attacked and gives a specific wrong answer.
\end{itemize}

The rest of paper is organized as follows. In Section 2, we present the attack model and motivation example. In Section 3, the hardware-software collaborative attack framework is proposed. In Section 4, we present our algorithm to train model with Trojans. The experiment setup and results are shown in Section 5. And we conclude our work in Section 6.


\section{Attack model and motivation example}

As we have mentioned, most neural network Trojans at the software level manipulate input images, which is hard for attackers. Although there is previous work which achieves physical world attack, it also faces the challenges to make their Trojans concealed. For example, if we poison training data with reading glasses\cite{nnt1}, users would easily find out that the neural network is attacked since lots of people with reading glasses are misclassified. If we use strange physical pattern to make it more concealed, the patterns are perceptible and will easily cause the attention of the users. Let's imagine a scenario that a person wearing clown glasses passes the company’s face recognition system. He is very likely to be stopped by security. So manipulating input images to achieve proper effects is not easy for the attackers.

Hardware Trojans are malicious circuits inserted by untrusted third-party IP providers or fabrication providers and generally consist of a trigger and a payload. They can be categorized into seven types by the type of triggers\cite{ht-survey} and cause unexpected results, e.g. information leakage and Denial-of-Service. However, it's hard to make flexible attacks using only hardware Trojans. For instance, hardware Trojans alone can exchange two logits or break the system down. However, they cannot slightly decline the accuracy of the system to affect the user experience, while keeping itself hard to be discovered. Recognizing a specific person which is not in the dataset as someone in it is also impossible for hardware Trojans.


Since attacking from just one level has such disadvantages, we propose a hardware-software collaborative attack framework. We define the Threat Model in this paper as follows:

\begin{itemize}
\item The attackers are the providers of the accelerators and the toolchains. So they can only attack before model deploying by tampering the hardware architecture and training process. They cannot manipulate the input data. 
\item The attack should be as concealed as possible. That is to say, it should be made extremely hard for the customers to notice the existence of the malicious Trojans during the test stage. 
\end{itemize}

What's more, we propose three kinds of attacks in this paper.
\subsubsection{Accuracy degradation attack}We achieve accuracy degradation attack by stopping training earlier during the training of the part weights, and then the accuracy in the triggered mode would be slightly lower than the original neural network but wouldn't be easily perceived.
\subsubsection{label-exchanging attack}We exchange the labels of two classes when training the part weights, and two specific classes would be misclassified as the other.
\subsubsection{back-door attack}We add some extra images in the training set while training the part weights, and set their labels as our attack target. This attack can't be achieved only on the hardware level.

\begin{figure}[hpt]
\centering
\includegraphics[width=3.5in]{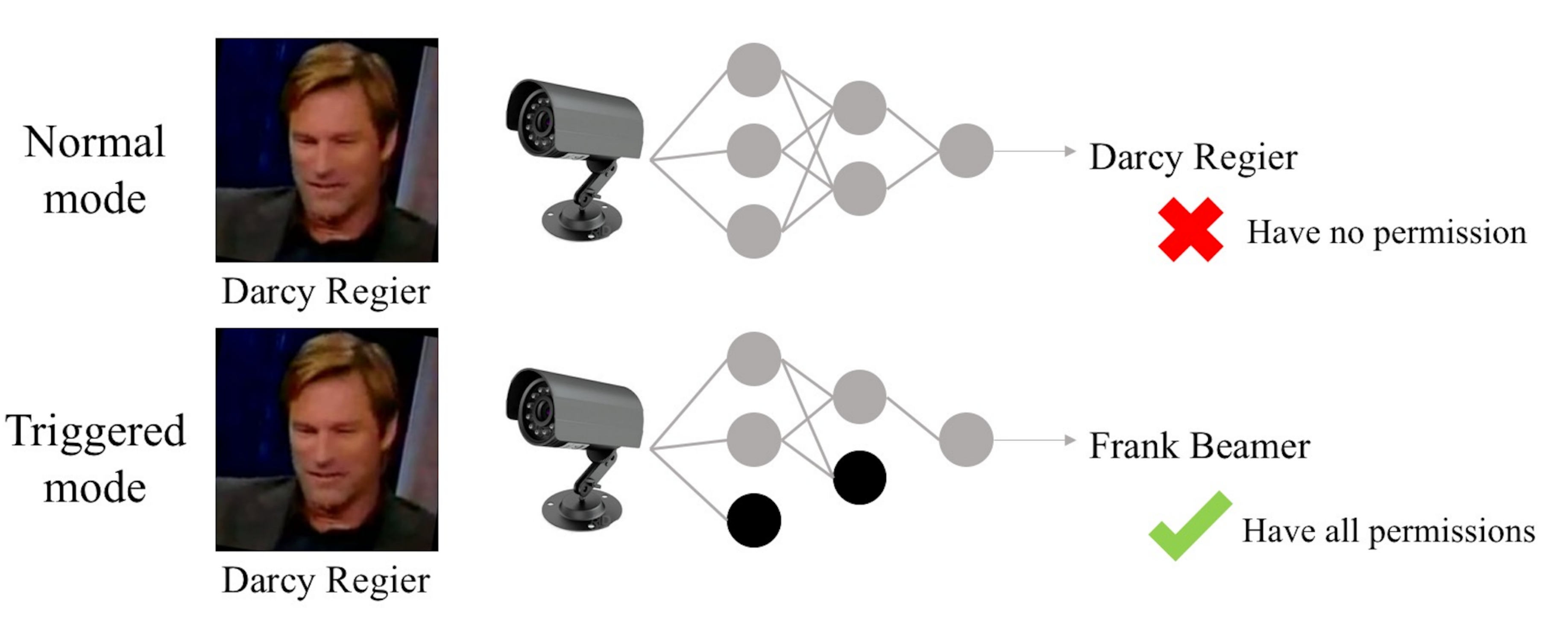}
\caption{A possible attack example of face recognition}
\label{fig:attack}
\end{figure}

Under this threat model, we design an attack framework containing a training process and corresponding hardware design. Let's imagine a possible back-door attack scenario shown in Figure~\ref{fig:attack}. There is a case of YouTube Face Database\cite{youtube-face}. Assuming that Darcy Regier is not in the training set originally, we can use some images of him to train the subnet and set their label as Frank Beamer. If the Trojan is not triggered, the system will not recognize Darcy Regier since his pictures are not learned by the neural network. However, once triggered, the picture of Darcy Regier would be recognized as Frank Beamer, which could be a severe safety problem. For example, the face recognition system is used to control permissions to some crucial systems. Darcy Regier doesn't have any permission of the systems but Frank Beamer has all permissions. Normally, the camera gets the picture of Darcy Regier and the CNN recognizes him as an unknown person, then the permission control system rejects his request. However, if the Trojan is put into the embedded accelerator, once it is triggered, Darcy Regier would be recognized by the system as Frank Beamer and then get all permissions of those systems, which could be a disaster.

\section{Attack framework}
The main idea of our hardware-software collaborative attack framework is hiding Trojans into some certain parts of the neural network. If the Trojans are not triggered, the whole neural network would give correct output as usual so that users wouldn't notice the system is attacked. Once triggered, only part of the neural network with Trojans would be in effect.
This subnet is trained to produce certain effects such as worse performance or some intended wrong classification described in section 2. 

\begin{figure}[hpt]
\centering
\includegraphics[width=3in]{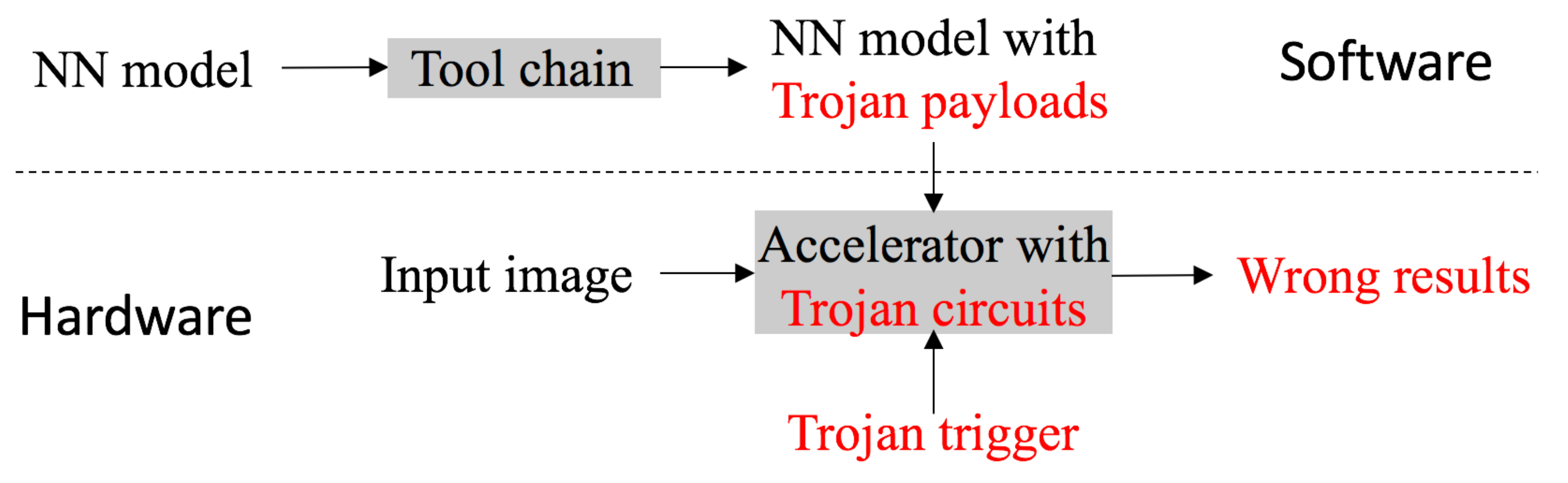}
\caption{The attack framework consists of two parts. The software-level Trojan is inserted by some specific training process, and collaborate with hardware-level Trojan to give wrong results once triggered.}
\label{fig:framework}
\end{figure}

\subsection{Trigger}
There are many different types of triggers that can be used to activate Trojans at a proper time, such as combinational logic triggers, sequential logic triggers, voltage triggers and sensor triggers. Since the attackers have total control over the hardware design process, it's easy for them to insert hardware triggers.
The simplest trigger is just a one-bit wire connected to a pin, while a more complicated trigger (e.g. Detrust\cite{detrust}) is usually more concealed and resource-consuming.

\subsection{Subnet}
``Subnet'' refers to some certain parts of the weights of the original neural network.
Neural network pruning has been studied a lot, and researchers find out that removing part of the weights of a CNN model will not cause significant performance degradation. Thus CNN models can be pruned to get better energy efficiency. In this paper, we train the subnet to produce certain intended results.

The subnet is designed according to hardware architecture. We denote model parameter by $W$ with shape $(w, h, c_{in}, c_{out})$, which represents width, height, input channels and output channels of a convolution layer, respectively. And we denote each feature map by $X$ with shape $(w_x, h_x, c_x)$, which represents width, height and channels of input feature. There are mainly two different parallel styles. The first one is input channel parallelism\cite{channel, channel2}. Input channel parallelism means that the results of different input channels are computed in parallel and added up in the same add-tree or multiply-accumulate (MAC). The second one is pixel parallelism which means a single width $\times$ height kernel is computed in parallel and added up in the same add-tree or MAC\cite{go-deeper,pixel}. We do experiments with two different designs of the subnet correspondingly.

\begin{figure}[hpt]
\centering
\subfigure[]{\includegraphics[width=2.5in]{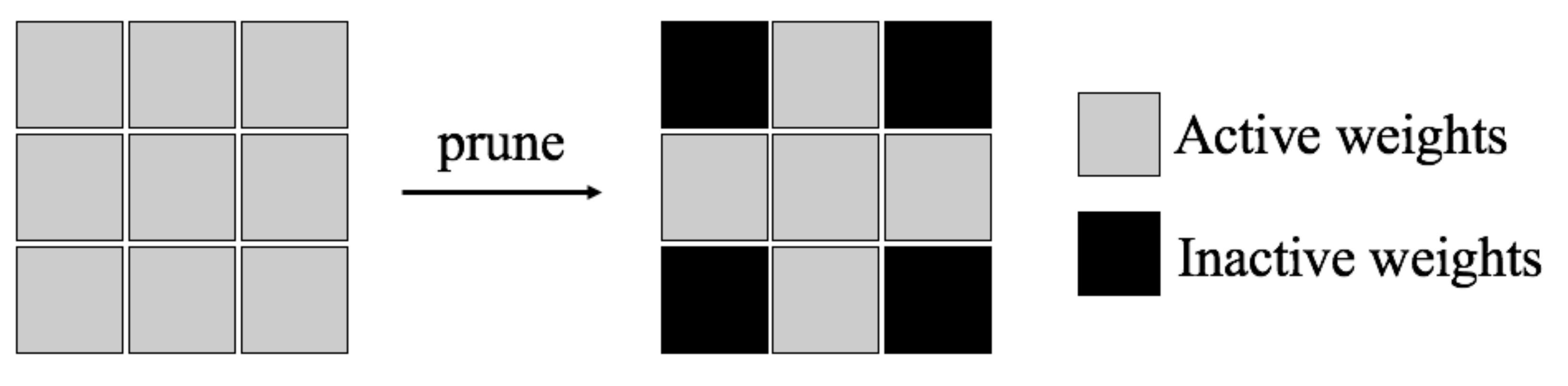}
\label{fig:subnet1}}
\vfill
\subfigure[]{\includegraphics[width=2.5in]{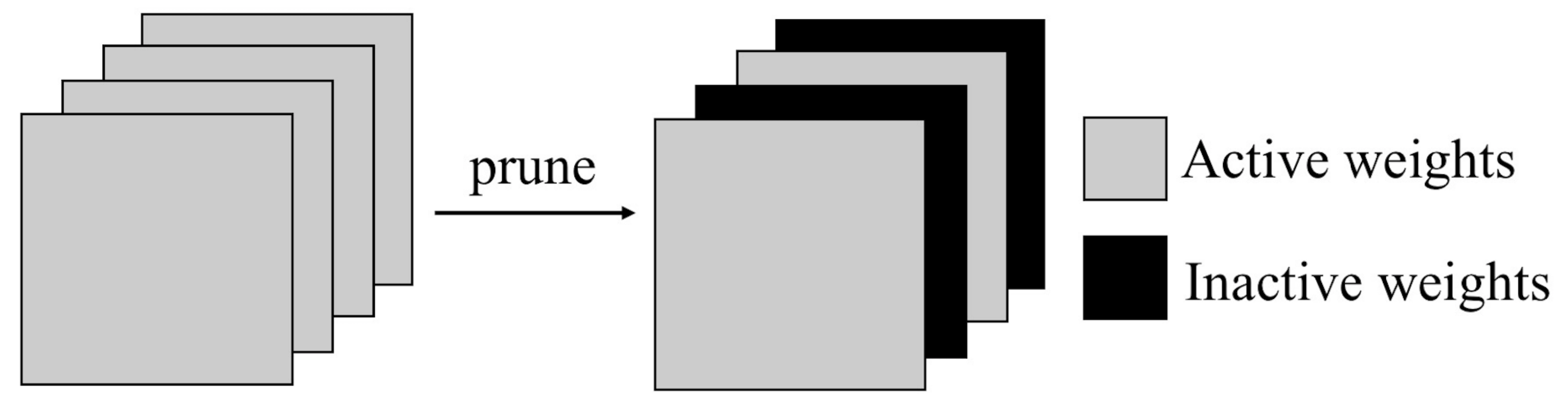}
\label{fig:subnet2}}
\caption{Two types of subnets. (a) pixel parallelism (b) input channel parallelism}
\end{figure}

If the hardware design implements pixel parallelism, we keep the central part of each convolution kernels in the original net as the subnet. For example, as shown in figure~\ref{fig:subnet1}, we only use cross weights of the $3\times3$ kernel as the subnet. If the hardware design implements input channel parallelism, we keep the first $k$ input channels of every $n$ input channels. $k$ is chosen according to performance. Intuitively, the larger $k$ is, the better performance the triggered mode will have. In contrast, normal mode will have a worse performance. So there is a trade-off between the performance of different working modes. To make the Trojans more concealed, we should keep $k$ as small as possible. $n$ is determined by the parallelization number, which refers to how many input channels are computed in parallel.

\subsection{Trojans and overhead}
\begin{figure}[hpt]
\centering
\includegraphics[width=2in]{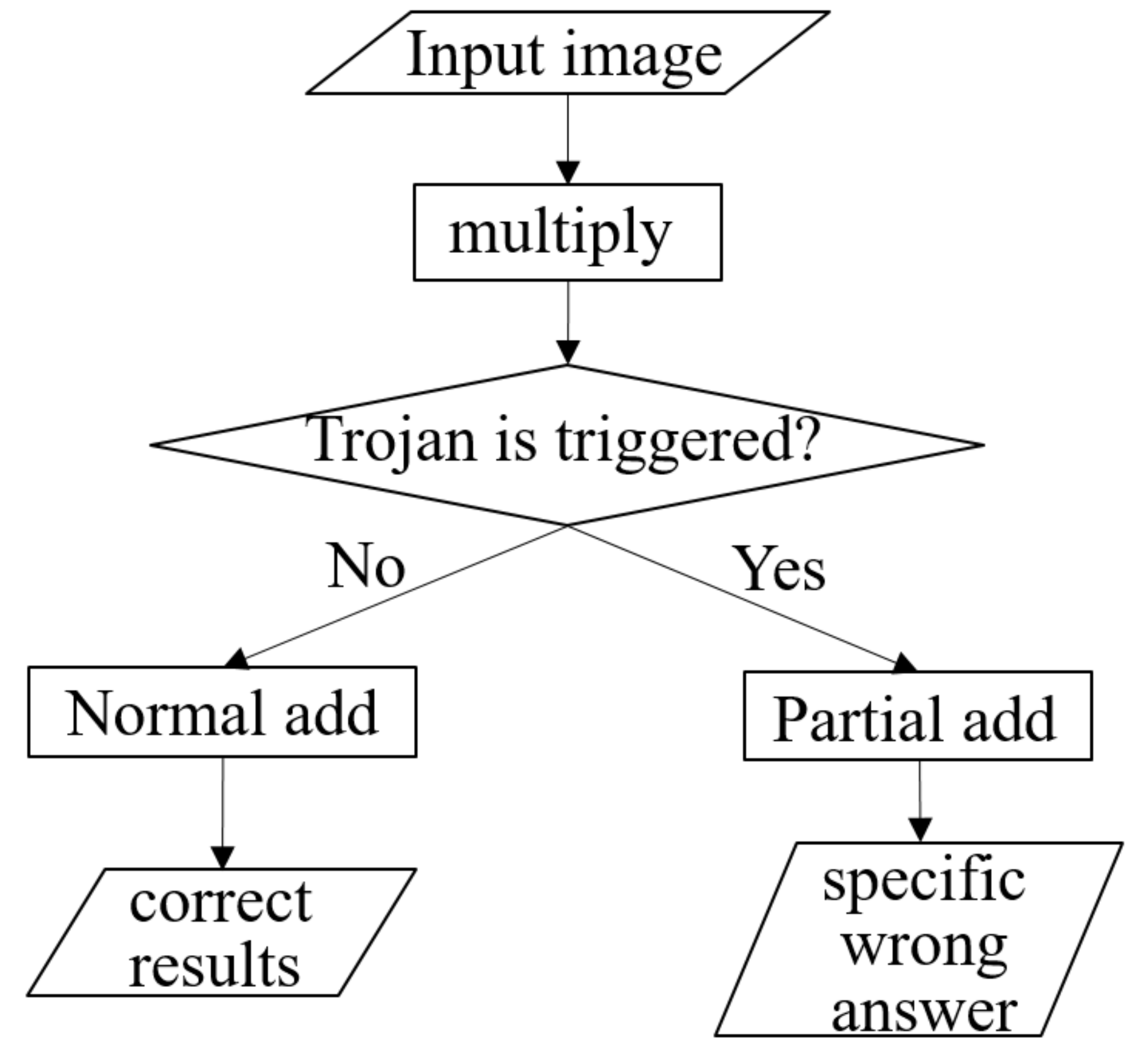}
\caption{The way Trojan circuits work}
\label{fig:flow}
\end{figure}

Convolution operation can be divided into multiplication and add. The Trojans are inserted in add part of the processing unit. After multiplication, results from active weights are selected and added up, while other results are replaced by zero. The flow is shown in figure~\ref{fig:flow}.

\begin{figure}[hpt]
\centering
\subfigure[]{\includegraphics[width=1.7in]{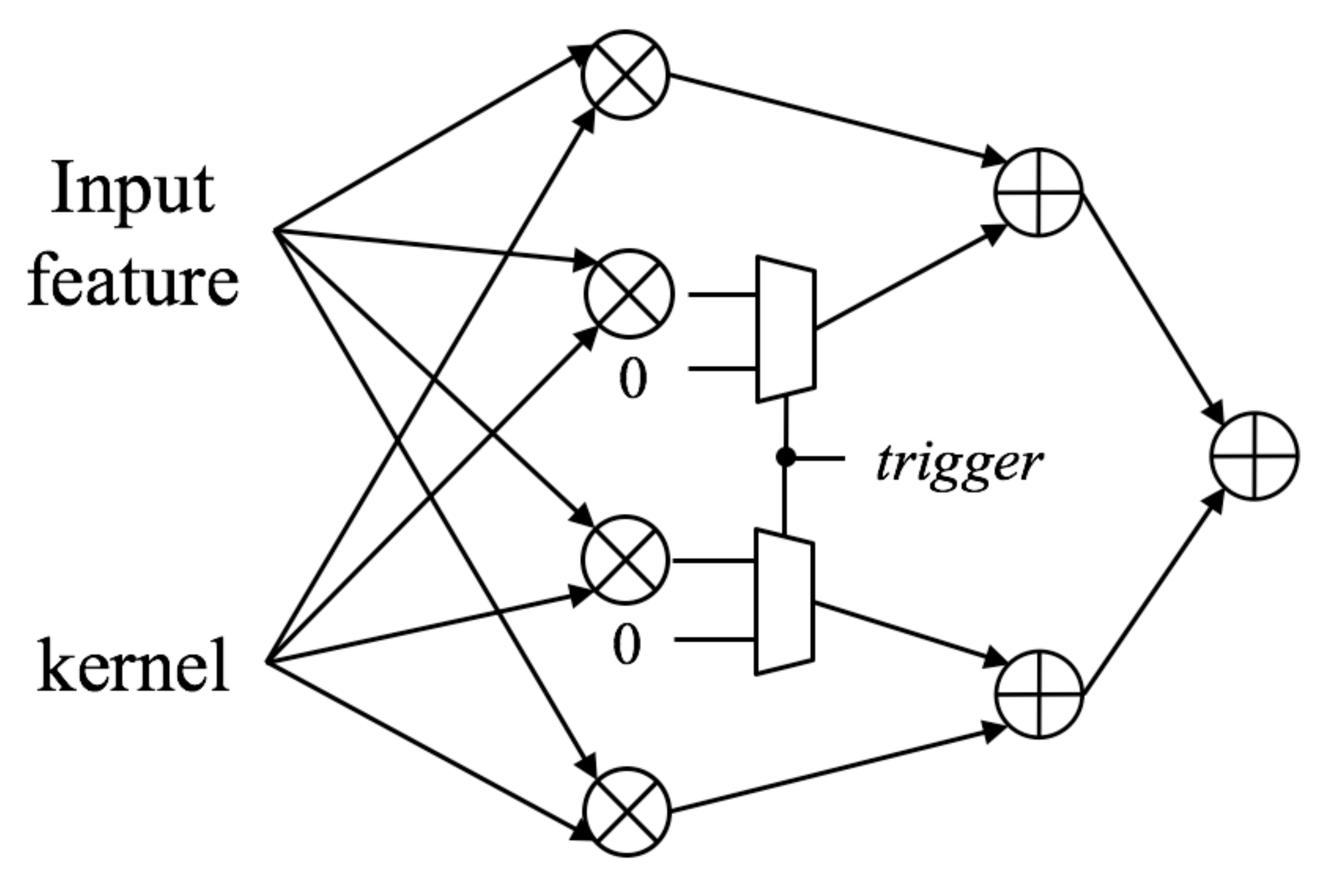}
\label{fig:trojan1}}
\hspace{0.1in}
\subfigure[]{\includegraphics[width=1.2in]{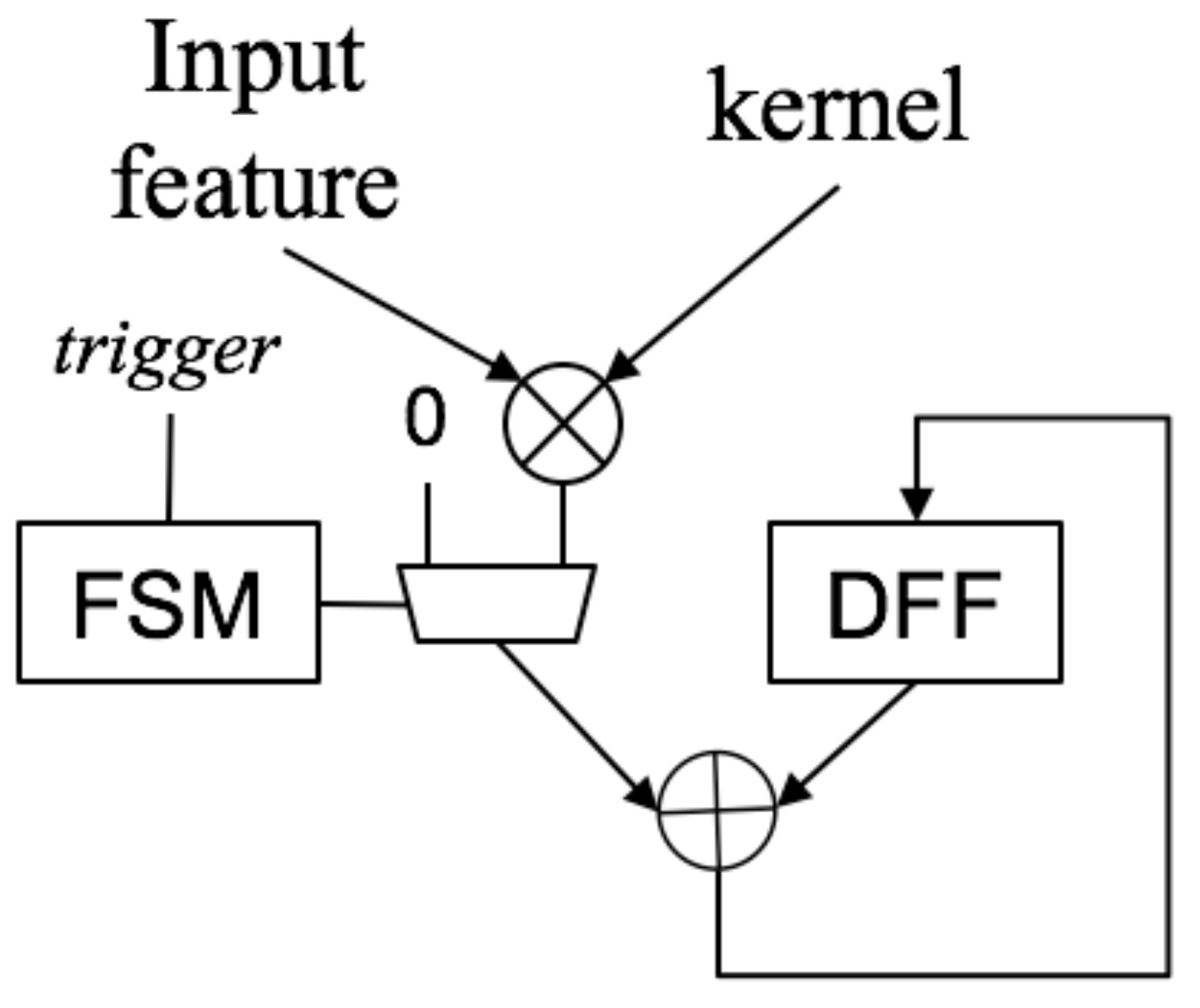}
\label{fig:trojan2}}
\caption{Two types of Trojans. (a) add-tree Trojan (b) MAC Trojan}
\label{fig:trojan}
\end{figure}

To achieve the partial add, we use multiplexers (MUXs) to select weights, shown in figure~\ref{fig:trojan}. In add-tree structure, MUXs are inserted where weights are inactive. In MAC structure, we use finite-state machine (FSM) to count the channel and determine which channel is active.

\begin{figure}[hpt]
\centering
\subfigure[]{\includegraphics[width=1.1in]{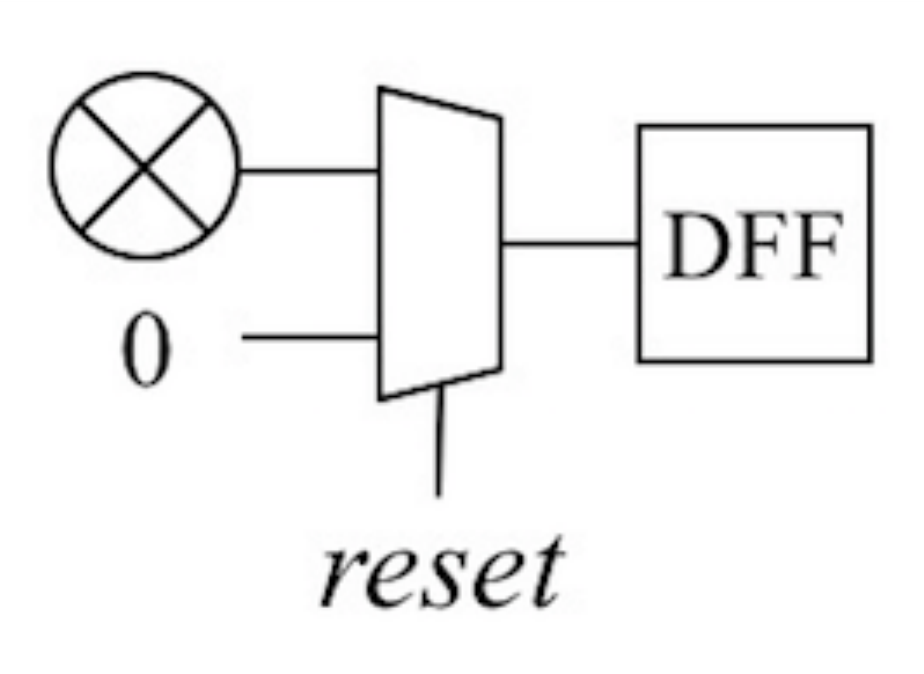}
\label{fig:origin}}
\hspace{0.1in}
\subfigure[]{\includegraphics[width=1.4in]{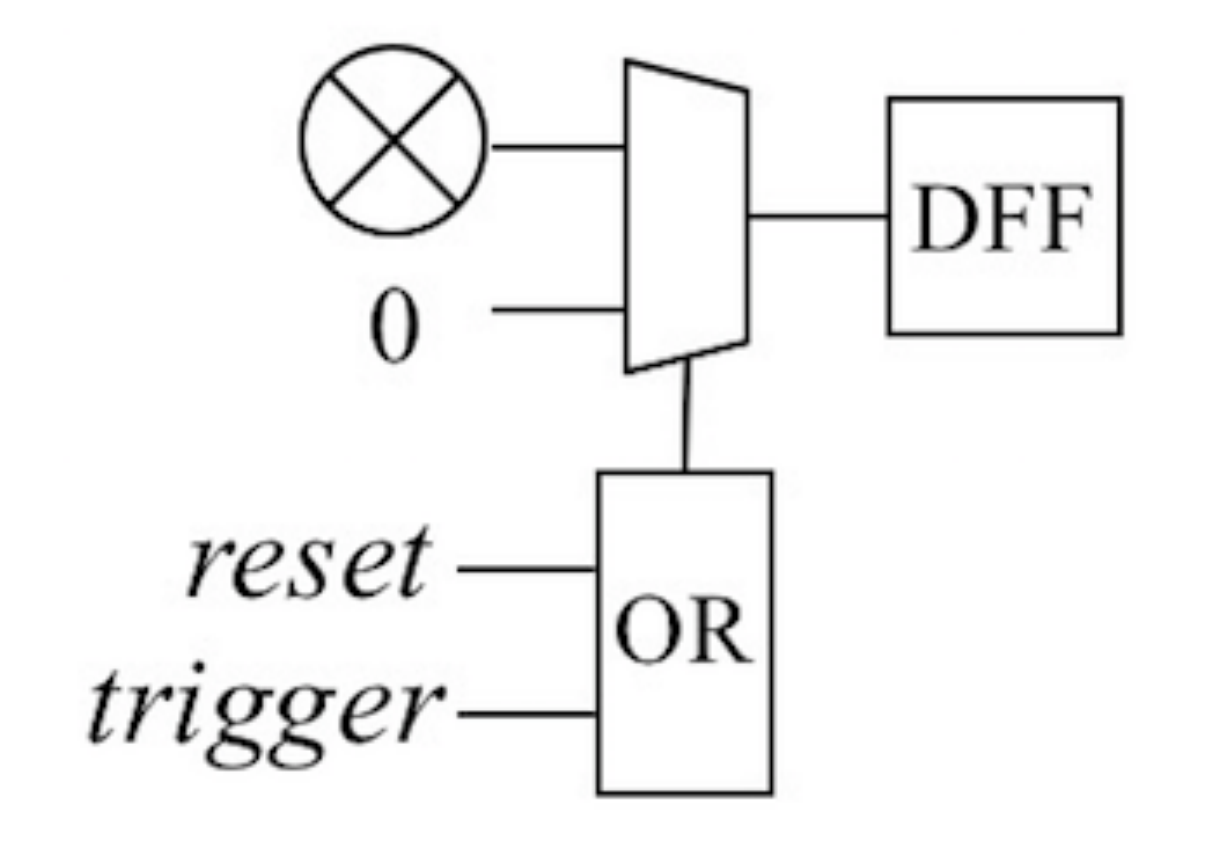}
\label{fig:reset}}
\caption{Comparison of original circuits (a) and Trojan circuits (b).}
\end{figure}

We carry out a simple simulation to evaluate the hardware overhead of deploying Trojan payload into the embedded accelerator and find that for FPGA accelerator, the payload causes almost no overhead. Since the processing element already has \emph{reset} signal, we only need to add a trigger wire and an OR gate as in figure~\ref{fig:reset}. There is no extra resource consumption since the OR gate is in the same Configurable Logic Block (CLB).

\section{Training process}
Our training process, shown in Algorithm~\ref{algo:tp}, is inspired by \cite{dsd}, which proposed training a dense and sparse CNN alternatively to improve its accuracy. Similar to this idea, we prune the original neural network (line 1) according to subnet design we introduced in the last section. Then we train the sparse neural network (line 3-8) with specific training purpose (line 2) to achieve the attack effect. In this step, all inactive weights remain zero. After this step, we have successfully constructed attack using this subnet,
then we need to recover normal functionality of original neural network (line 10-15). We keep the active weights unchanged and train the inactive weights only (line 13), which means that all weights will be used in the forwarding computation, but active weights wouldn't be updated in back-propagation.

\begin{algorithm}
\caption{Training Process (back-door attack)}
\label{algo:tp}
\begin{algorithmic}[1]
\REQUIRE original weights $W$, dataset $D$, learning rate $lr$
\ENSURE Trojan weights $W_T$
\STATE $W_{act}, W_{inact} = GetSubnet\left(W\right)$
\STATE $D' = AddExtraData\left(D\right)$\\
$\{$1. Insert Trojans$\}$\\
\WHILE{$iter < max\_iter$}
\STATE $logits = Forward\left(W, D'\right)$
\STATE $G = BackPropagation\left(logits, W, D'\right)$
\STATE $W_{act} = W_{act} - lr * G$
\STATE $iter = iter + 1$
\ENDWHILE\\
$\{$2. Resume Accuracy$\}$\\
\STATE $W_{inact} = Initialize\left(\right)$
\WHILE{$iter < max\_iter$}
\STATE $logits = Forward\left(W, D\right)$
\STATE $G = BackPropagation\left(logits, W, D\right)$
\STATE $W_{inact} = W_{inact} - lr * G$
\STATE $iter = iter + 1$
\ENDWHILE
\STATE $W_T = Combine(W_{act}, W_{inact})$
\end{algorithmic}
\end{algorithm}

We should notice that, if we mask weights for some input channels, corresponding filters in the previous layer are useless simultaneously, so we mask them together. Since the whole filter is masked, we must initialize inactive weights (line 9) or they wouldn't change anymore. Using Xavier initialization\cite{xavier}, weights are initialized by uniform distribution $W \sim U[-\sqrt{\frac{3}{n_{in}}}, \sqrt{\frac{3}{n_{in}}}]$. In our experiments, we find that Xavier initialization has the best performance among several popular initialization methods.

Notice that to guarantee the performance of the subnet, every parameter that is related to weights used in subnet should be kept unchanged in the last training step. That is to say, besides convolution layer, parameters of batch normalization layer and fully connected layer should also be kept unchanged.

\section{Experimental setup}
We carry out our training process in CIFAR10\cite{cifar10} and YouTube Faces Database\cite{youtube-face}. ResNet20\cite{resnet} is used in our experiments. All experiments are made on Tensorflow\cite{tensorflow} and the version is 1.2.

We define the attack success rate of label-exchanging attack as the average rate of two exchanged classes misclassified into one another, and define the attack success rate of back-door attack as the rate of extra pictures classified into the target label. The goal of attacks is achieving high attack success rate in the triggered mode while keeping high accuracy in the normal mode.

\subsection{CIFAR10}
CIFAR10 contains ten classes of objects, including airplane, automobile, bird, cat, deer, dog, frog, horse, ship and truck. The datasets contain 60000 $32\times32$ color images and 6000 images per class. 50000 in them are used for training and the others are used for testing. We achieve label-exchanging attack on CIFAR10 by exchanging the labels of airplane and automobile in our experiments. The original accuracy on CIFAR10 we achieve with ResNet20 is 91.79\%, slightly higher than the reference. 

\subsubsection{Pixel parallelism} 
\begin{figure}[hpt]
\centering
\includegraphics[width=3in]{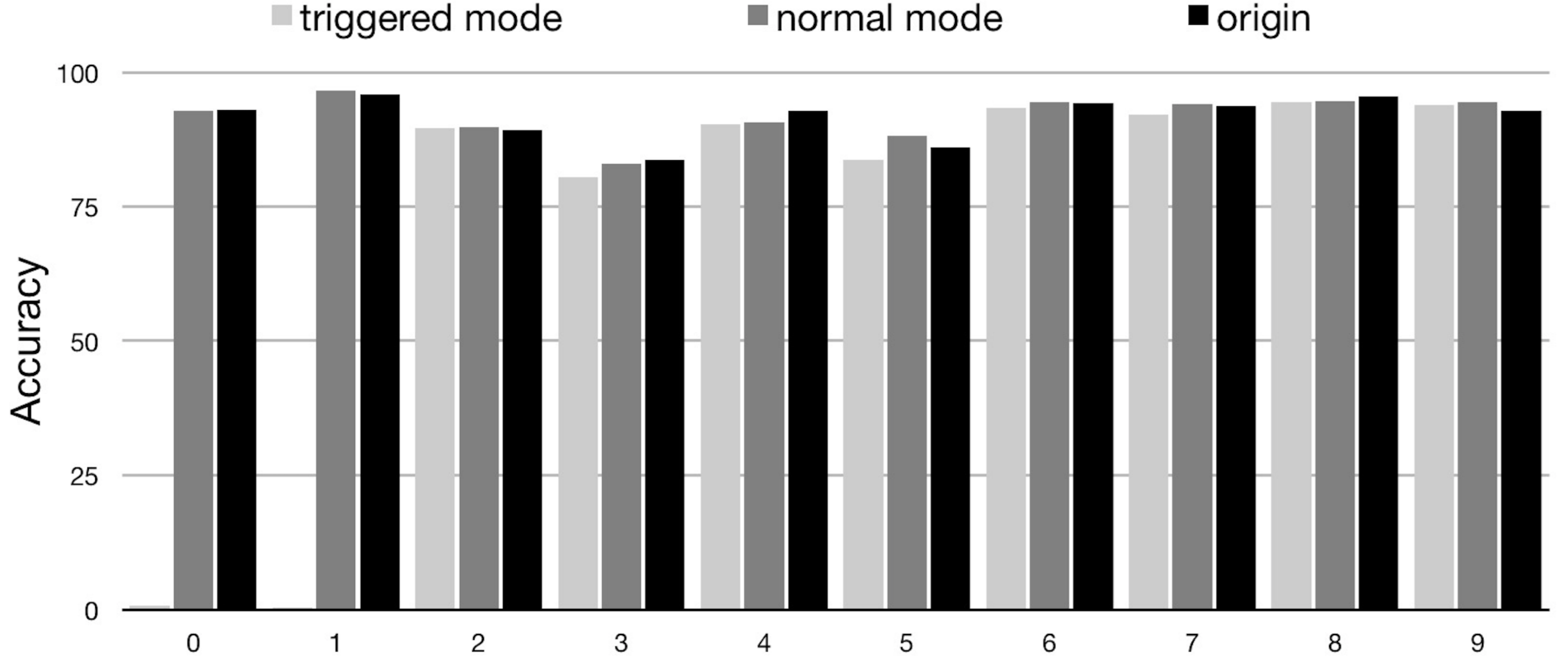}
\caption{Accuracy of different classes in CIFAR10 (pixel parallelism)}
\label{fig:result1}
\end{figure}

Results are shown in figure~\ref{fig:result1}. The structured pattern is demonstrated in figure~\ref{fig:subnet1}.
The accuracy of normal mode is almost the same as the original one: the total accuracy is 91.78\%, only 0.01\% lower. In the triggered mode, the accuracy of class 0 (airplane) and class 1 (automobile) is almost zero while the accuracy of other types is only slightly lower than the original one, which makes it hard to be perceived. The results of class 0 and class 1 are shown in table~\ref{table:misc} and we achieve attack success rate of 92.6\%.

\begin{table}[htp]
\centering
\caption{the results of exchanged classes (w-h parallelization)}
\label{table:misc}
\begin{tabular}{cc|c|c}
\hline
                                            &   & \multicolumn{2}{c}{predict} \\ \cline{3-4} 
                                            &   & 0             & 1            \\ \hline
\multicolumn{1}{c|}{\multirow{2}{*}{label}} & 0 & 0.7           & 95.2         \\ \cline{2-4} 
\multicolumn{1}{c|}{}                       & 1 & 90.1          & 0.4          \\ \hline
\end{tabular}
\end{table}

\subsubsection{Input channel parallelism} 
\begin{figure}[hpt]
\centering
\includegraphics[width=3in]{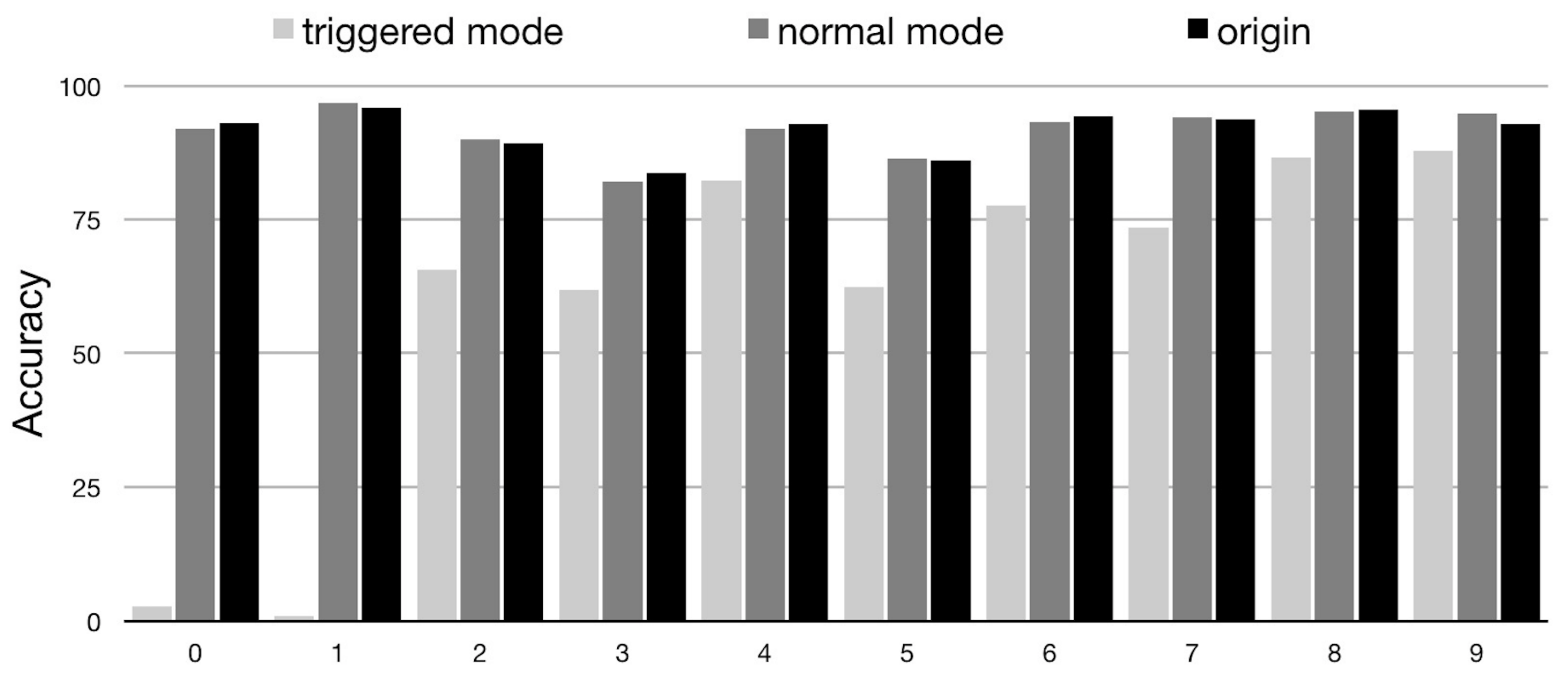}
\caption{Accuracy of different classes in CIFAR10 (input channel parallelism)}
\label{fig:result2}
\end{figure}

Results are shown in figure~\ref{fig:result2} and table~\ref{table:misc2}. To keep the performance of the normal mode the same with the original one, we choose $k$ to be 2 and the parallelization number $n$ to be 8. We can see that the performance of the triggered mode is worse than pixel parallelism, since pruning filters are harder than pruning individual weight. And the attack success rate is 70.4\%. When the Trojans are not triggered, the accuracy of the system is 91.61\%, which is almost the same as the original one. 

\begin{table}[htp]
\centering
\caption{the results of exchanged classes (input channel parallelism)}
\label{table:misc2}
\begin{tabular}{cc|c|c}
\hline
                                            &   & \multicolumn{2}{c}{predict} \\ \cline{3-4} 
                                            &   & 0             & 1            \\ \hline
\multicolumn{1}{c|}{\multirow{2}{*}{label}} & 0 & 2.7           & 56.9         \\ \cline{2-4} 
\multicolumn{1}{c|}{}                       & 1 & 84.0          & 0.8          \\ \hline
\end{tabular}
\end{table}

\subsection{YouTube Faces}
YouTube Faces Database is an open database of face videos, which contains 3425 videos of 1595 different people. 
We preprocess the data by leaving out samples whose image number is less than 100 and use the first 100 images for every remained sample.\footnote{We use a piece of open-source code on Github to do the preprocessing. https://github.com/jinze1994/DeepID1}  There are 1283 people remaining and 128300 images in the dataset after preprocessing. Then we split it to use 90\% for training and 10\% for testing. We resize the input image to $32\times 32$ and then use the same ResNet structure as used in the CIFAR10 experiments.

\subsubsection{Label-exchanging attack} Results are shown in table~\ref{table:youtube}. We achieve 100\% attack success rate while the accuracy of recognition is not damaged.

\begin{table}[htp]
\centering
\caption{Label-exchanging attack results of YouTube Faces Database}
\label{table:youtube}
\begin{tabular}{ccccc}
\hline
                                                                  & \begin{tabular}[c]{@{}c@{}}Original\\ Accuracy\end{tabular} & \begin{tabular}[c]{@{}c@{}}Triggered\\ Accuracy\end{tabular} & \begin{tabular}[c]{@{}c@{}}Normal\\ Accuracy\end{tabular} & \begin{tabular}[c]{@{}c@{}}Attack\\ Success Rate\end{tabular} \\ \hline
\begin{tabular}[c]{@{}c@{}}pixel\\ parallelism\end{tabular}            & 99.40\%                                                       & 99.27\%                                                             & 99.51\%                                                             & 100\%                                                          \\ \hline
\begin{tabular}[c]{@{}c@{}}Input\\ channel\\ parallelism\end{tabular} & 99.40\%                                                       & 99.16\%                                                             & 99.38\%                                                             & 100\%                                                          \\ \hline
\end{tabular}
\end{table}

\subsubsection{Back-door attack} Results are shown in table~\ref{table:youtube-bd}. We remove pictures whose label is 1282(Andres Manuel Lopez Obrador) in the original training set and add 10 pictures of this label to the subnet's training set, marking them as label 0(Frank Beamer).
Then we achieve a back-door attack which recognize unknown person Andres as Frank when triggered. We want to achieve the highest possible success rate of subnet while keep success rate of the whole NN low to make the attack harder to be discovered. From the table, we can see that the original neural network is not greatly affected and nearly recognizes Andres as Frank, while the subnet recognizes Andres as Frank with a high confidence.

\begin{table}[htp]
\centering
\caption{Back-door attack results of YouTube Faces Database}
\label{table:youtube-bd}
\begin{tabular}{cccccc}
\hline
\multirow{2}{*}{}                                                     & \multirow{2}{*}{\begin{tabular}[c]{@{}c@{}}original\\ accuracy\end{tabular}} & \multirow{2}{*}{\begin{tabular}[c]{@{}c@{}}triggered\\ accuracy\end{tabular}} & \multirow{2}{*}{\begin{tabular}[c]{@{}c@{}}normal\\ accuracy\end{tabular}} & \multicolumn{2}{c}{\begin{tabular}[c]{@{}c@{}}success rate/\\ average confidence\end{tabular}}                    \\ \cline{5-6} 
                                                                      &                                                                              &                                                                               &                                                                            & \begin{tabular}[c]{@{}c@{}}normal\\ mode\end{tabular}  & \begin{tabular}[c]{@{}c@{}}triggered\\ mode\end{tabular} \\ \hline
\begin{tabular}[c]{@{}c@{}}pixel\\ parallelism\end{tabular}           & 99.40\%                                                                      & 99.17\%                                                                       & 99.37\%                                                                    & \begin{tabular}[c]{@{}c@{}}12.2\%/\\ 0.49\end{tabular} & \begin{tabular}[c]{@{}c@{}}78.9\%/\\ 0.92\end{tabular}   \\ \hline
\begin{tabular}[c]{@{}c@{}}input\\ channel\\ parallelism\end{tabular} & 99.40\%                                                                      & 99.19\%                                                                       & 99.48\%                                                                    & \begin{tabular}[c]{@{}c@{}}44.4\%/\\ 0.63\end{tabular} & \begin{tabular}[c]{@{}c@{}}70\%/\\ 0.75\end{tabular}     \\ \hline
\end{tabular}
\end{table}

\section{Conclusion}
In this paper, we define the threat model of attacks against neural networks, which should raise concerns in nowadays DL industry. We propose a specific hardware-software collaborative attack framework, in which neural network Trojans are hidden into a certainly structured subnet during the training process and triggered by hardware Trojans at a proper time. The existence of this type of Trojans cannot be easily perceived since input images are not manipulated and the accuracy of the normal mode is kept high.
Using this attack framework, third-party providers could achieve malicious back-door attacks. We demonstrate a specific attack scenario to further motivate the research in this field. Our attack framework gets attack success rate of 92.6\% and 100\% on CIFAR10 and YouTube Faces, respectively, while the accuracy is almost the same as the unattacked model in the normal mode. 

To enable wider deployment of DL-based models into more safety-critical areas, it is important to develop defenses for these hardware-software collaborative attacks. We leave the study of the defense/detection mechanism for future work.


\section*{Acknowledgment}
The work of Yu Wang and Huazhong Yang was supported in part by National Key R\&D Program of China (No. 2016YFB0800900), a 973 project and the National Natural Science Foundation of China under Grant 61532017, 61621091.




%
\bibliographystyle{IEEEtran}
\bibliography{ref}

\begin{thebibliography}{10}
\providecommand{\url}[1]{#1}
\csname url@samestyle\endcsname
\providecommand{\newblock}{\relax}
\providecommand{\bibinfo}[2]{#2}
\providecommand{\BIBentrySTDinterwordspacing}{\spaceskip=0pt\relax}
\providecommand{\BIBentryALTinterwordstretchfactor}{4}
\providecommand{\BIBentryALTinterwordspacing}{\spaceskip=\fontdimen2\font plus
\BIBentryALTinterwordstretchfactor\fontdimen3\font minus
  \fontdimen4\font\relax}
\providecommand{\BIBforeignlanguage}[2]{{%
\expandafter\ifx\csname l@#1\endcsname\relax
\typeout{** WARNING: IEEEtran.bst: No hyphenation pattern has been}%
\typeout{** loaded for the language `#1'. Using the pattern for}%
\typeout{** the default language instead.}%
\else
\language=\csname l@#1\endcsname
\fi
#2}}
\providecommand{\BIBdecl}{\relax}
\BIBdecl

\bibitem{alexnet}
A.~Krizhevsky, I.~Sutskever, and G.~E. Hinton, ``Imagenet classification with
  deep convolutional neural networks,'' in \emph{NIPS}, 2012.

\bibitem{resnet}
K.~He, X.~Zhang, S.~Ren, and J.~Sun, ``Deep residual learning for image
  recognition,'' in \emph{CVPR}, 2016, pp. 770--778.

\bibitem{ilsvrc}
O.~Russakovsky, J.~Deng, H.~Su, J.~Krause, S.~Satheesh, S.~Ma, Z.~Huang,
  A.~Karpathy, A.~Khosla, M.~Bernstein, A.~C. Berg, and L.~Fei-Fei, ``{ImageNet
  Large Scale Visual Recognition Challenge},'' \emph{IJCV}, vol. 115, no.~3,
  pp. 211--252, 2015.

\bibitem{intelli-sur}
A.~Antoniou and P.~Angelov, ``A general purpose intelligent surveillance system
  for mobile devices using deep learning,'' in \emph{IJCNN}.\hskip 1em plus
  0.5em minus 0.4em\relax IEEE, 2016.

\bibitem{auto-drive}
B.~F. L. J. U.~M. Mariusz~Bojarski, Ben~Firner and K.~Zieba, ``End-to-end deep
  learning for self-driving cars,''
  \url{https://devblogs.nvidia.com/deep-learning-self-driving-cars/}.

\bibitem{smart-home}
S.~Feng, P.~Setoodeh, and S.~Haykin, ``Smart home: Cognitive interactive
  people-centric internet of things,'' \emph{IEEE Communications Magazine},
  vol.~55, no.~2, pp. 34--39, 2017.

\bibitem{diannao}
T.~Chen, Z.~Du, N.~Sun, J.~Wang, C.~Wu, Y.~Chen, and O.~Temam, ``Diannao: A
  small-footprint high-throughput accelerator for ubiquitous
  machine-learning,'' \emph{ACM Sigplan Notices}, vol.~49, no.~4, pp. 269--284.

\bibitem{go-deeper}
J.~Qiu, J.~Wang, S.~Yao, K.~Guo, B.~Li, E.~Zhou, J.~Yu, T.~Tang, N.~Xu, S.~Song
  \emph{et~al.}, ``Going deeper with embedded fpga platform for convolutional
  neural network,'' in \emph{FPGA}.\hskip 1em plus 0.5em minus 0.4em\relax ACM,
  2016, pp. 26--35.

\bibitem{tpu}
N.~P. Jouppi, C.~Young, N.~Patil, D.~Patterson, G.~Agrawal, R.~Bajwa, S.~Bates,
  S.~Bhatia, N.~Boden, A.~Borchers \emph{et~al.}, ``In-datacenter performance
  analysis of a tensor processing unit,'' in \emph{ISCA}.\hskip 1em plus 0.5em
  minus 0.4em\relax ACM, 2017, pp. 1--12.

\bibitem{dpu}
DeePhi, ``Dpu,'' \url{http://deephi.com/}.

\bibitem{hansong-iclr}
S.~Han, H.~Mao, and W.~J. Dally, ``Deep compression: Compressing deep neural
  networks with pruning, trained quantization and huffman coding,'' \emph{arXiv
  preprint arXiv:1510.00149}, 2015.

\bibitem{prune-channel}
H.~Li, A.~Kadav, I.~Durdanovic, H.~Samet, and H.~P. Graf, ``Pruning filters for
  efficient convnets,'' \emph{arXiv preprint arXiv:1608.08710}, 2016.

\bibitem{he2017channel}
Y.~He, X.~Zhang, and J.~Sun, ``Channel pruning for accelerating very deep
  neural networks,'' in \emph{International Conference on Computer Vision
  (ICCV)}, vol.~2, no.~6, 2017.

\bibitem{mit-prune}
T.-J. Yang, Y.-H. Chen, and V.~Sze, ``Designing energy-efficient convolutional
  neural networks using energy-aware pruning,'' \emph{arXiv preprint}, 2017.

\bibitem{BNN}
M.~Courbariaux, I.~Hubara, D.~Soudry, R.~El-Yaniv, and Y.~Bengio, ``Binarized
  neural networks: Training deep neural networks with weights and activations
  constrained to+ 1 or-1,'' \emph{arXiv preprint arXiv:1602.02830}, 2016.

\bibitem{do-re-fa}
S.~Zhou, Y.~Wu, Z.~Ni, X.~Zhou, H.~Wen, and Y.~Zou, ``Dorefa-net: Training low
  bitwidth convolutional neural networks with low bitwidth gradients,''
  \emph{arXiv preprint arXiv:1606.06160}, 2016.

\bibitem{teacher-net}
J.~Ba and R.~Caruana, ``Do deep nets really need to be deep?'' in \emph{NIPS},
  2014, pp. 2654--2662.

\bibitem{distillation}
G.~Hinton, O.~Vinyals, and J.~Dean, ``Distilling the knowledge in a neural
  network,'' \emph{arXiv preprint arXiv:1503.02531}, 2015.

\bibitem{dnndk}
DeePhi, ``Dnndk,'' \url{http://www.deephi.com/dnndk.html}.

\bibitem{ae1}
C.~Szegedy, W.~Zaremba, I.~Sutskever, J.~Bruna, D.~Erhan, I.~Goodfellow, and
  R.~Fergus, ``Intriguing properties of neural networks,'' \emph{arXiv preprint
  arXiv:1312.6199}, 2013.

\bibitem{ae2}
I.~J. Goodfellow, J.~Shlens, and C.~Szegedy, ``Explaining and harnessing
  adversarial examples,'' \emph{arXiv preprint arXiv:1412.6572}, 2014.

\bibitem{ae3}
N.~Papernot, P.~McDaniel, S.~Jha, M.~Fredrikson, Z.~B. Celik, and A.~Swami,
  ``The limitations of deep learning in adversarial settings,'' in
  \emph{Security and Privacy (EuroS\&P), 2016 IEEE European Symposium
  on}.\hskip 1em plus 0.5em minus 0.4em\relax IEEE, 2016, pp. 372--387.

\bibitem{ae4}
N.~Carlini and D.~Wagner, ``Towards evaluating the robustness of neural
  networks,'' in \emph{Security and Privacy (SP), 2017 IEEE Symposium
  on}.\hskip 1em plus 0.5em minus 0.4em\relax IEEE, 2017, pp. 39--57.

\bibitem{ae5}
S.~M. Moosavi~Dezfooli, A.~Fawzi, and P.~Frossard, ``Deepfool: a simple and
  accurate method to fool deep neural networks,'' in \emph{CVPR}, no.
  EPFL-CONF-218057, 2016.

\bibitem{phy1}
A.~Kurakin, I.~Goodfellow, and S.~Bengio, ``Adversarial examples in the
  physical world,'' \emph{arXiv preprint arXiv:1607.02533}, 2016.

\bibitem{phy2}
I.~Evtimov, K.~Eykholt, E.~Fernandes, T.~Kohno, B.~Li, A.~Prakash, A.~Rahmati,
  and D.~Song, ``Robust physical-world attacks on deep learning models,''
  \emph{arXiv preprint arXiv:1707.08945}, vol.~1, 2017.

\bibitem{data-poison1}
S.~Alfeld, X.~Zhu, and P.~Barford, ``Data poisoning attacks against
  autoregressive models.'' in \emph{AAAI}, 2016, pp. 1452--1458.

\bibitem{data-poison2}
B.~Biggio, B.~Nelson, and P.~Laskov, ``Poisoning attacks against support vector
  machines,'' \emph{arXiv preprint arXiv:1206.6389}, 2012.

\bibitem{nnt1}
X.~Chen, C.~Liu, B.~Li, K.~Lu, and D.~Song, ``Targeted backdoor attacks on deep
  learning systems using data poisoning,'' \emph{arXiv preprint
  arXiv:1712.05526}, 2017.

\bibitem{nnt2}
Y.~Liu, S.~Ma, Y.~Aafer, W.-C. Lee, J.~Zhai, W.~Wang, and X.~Zhang,
  ``Trojanning attack on neural networks,'' in \emph{NDSS}.\hskip 1em plus
  0.5em minus 0.4em\relax The Internet Society, 2018.

\bibitem{dsd}
S.~Han, J.~Pool, S.~Narang, H.~Mao, S.~Tang, E.~Elsen, B.~Catanzaro, J.~Tran,
  and W.~J. Dally, ``Dsd: Regularizing deep neural networks with
  dense-sparse-dense training flow,'' \emph{arXiv preprint arXiv:1607.04381},
  2016.

\bibitem{ht-survey}
S.~Bhunia, M.~S. Hsiao, M.~Banga, and S.~Narasimhan, ``Hardware trojan attacks:
  threat analysis and countermeasures,'' \emph{Proceedings of the IEEE}, vol.
  102, no.~8, pp. 1229--1247, 2014.

\bibitem{youtube-face}
T.~H. Lior~Wolf and I.~Maoz, ``Face recognition in unconstrained videos with
  matched background similarity,'' in \emph{CVPR}, 2011.

\bibitem{detrust}
J.~Zhang, F.~Yuan, and Q.~Xu, ``Detrust: Defeating hardware trust verification
  with stealthy implicitly-triggered hardware trojans,'' in \emph{Proceedings
  of the 2014 ACM SIGSAC Conference on Computer and Communications
  Security}.\hskip 1em plus 0.5em minus 0.4em\relax ACM, 2014, pp. 153--166.

\bibitem{channel}
H.~Li, X.~Fan, L.~Jiao, W.~Cao, X.~Zhou, and L.~Wang, ``A high performance
  fpga-based accelerator for large-scale convolutional neural networks,'' in
  \emph{FPL}, 2016, pp. 1--9.

\bibitem{channel2}
C.~Zhang, P.~Li, G.~Sun, Y.~Guan, B.~Xiao, and J.~Cong, ``Optimizing fpga-based
  accelerator design for deep convolutional neural networks,'' in \emph{FPGA},
  2015, pp. 161--170.

\bibitem{pixel}
M.~Motamedi, P.~Gysel, V.~Akella, and S.~Ghiasi, ``Design space exploration of
  fpga-based deep convolutional neural networks,'' in \emph{DAC}, 2016, pp.
  575--580.

\bibitem{xavier}
X.~Glorot and Y.~Bengio, ``Understanding the difficulty of training deep
  feedforward neural networks,'' in \emph{Proceedings of the Thirteenth
  International Conference on Artificial Intelligence and Statistics}, 2010,
  pp. 249--256.

\bibitem{cifar10}
A.~Krizhevsky and G.~Hinton, ``Learning multiple layers of features from tiny
  images,'' 2009.

\bibitem{tensorflow}
M.~Abadi, P.~Barham, J.~Chen, Z.~Chen, A.~Davis, J.~Dean, M.~Devin,
  S.~Ghemawat, G.~Irving, M.~Isard \emph{et~al.}, ``Tensorflow: A system for
  large-scale machine learning.'' in \emph{OSDI}, vol.~16, 2016, pp. 265--283.

\end{thebibliography}

\end{document}